\documentclass[twocolumn, floatfix,prl,showpacs,amsmath,amssymb,superscriptaddress]{revtex4}

\usepackage[latin1]{inputenc}
\usepackage[T1]{fontenc}
\usepackage{amsmath}
\usepackage{amssymb,amsfonts,textcomp}
\usepackage{color}
\usepackage{array}
\usepackage{hhline}
\usepackage{graphicx}
\usepackage{hyperref}
\hypersetup{pdftex, colorlinks=true, linkcolor=blue, citecolor=blue, filecolor=blue, urlcolor=blue}
\newcommand\textsubscript[1]{\ensuremath{{}_{\text{#1}}}}
\newcommand\normalsubformula[1]{\text{\mathversion{normal}$#1$}}
\begin{document}
\title{Spectroscopy of spin-orbit quantum bits in indium antimonide nanowires}
\author{S. Nadj-Perge}
\affiliation{Kavli Institute of Nanoscience, Delft University of
Technology, 2600 GA Delft, The Netherlands}

\author{V. S. Pribiag}
\affiliation{Kavli Institute of Nanoscience, Delft University of
Technology, 2600 GA Delft, The Netherlands}

\author{J. W. G. van den Berg}
\affiliation{Kavli Institute of Nanoscience, Delft University of
Technology, 2600 GA Delft, The Netherlands}

\author{K. Zuo}
\affiliation{Kavli Institute of Nanoscience, Delft University of
Technology, 2600 GA Delft, The Netherlands}

\author{S. R. Plissard}
\affiliation{Department of Applied Physics, Eindhoven University
of Technology, 5600 MB Eindhoven, The Netherlands}

\author{E. P. A. M. Bakkers}
\affiliation{Kavli Institute of Nanoscience, Delft University of
Technology, 2600 GA Delft, The Netherlands}
\affiliation{Department of Applied Physics, Eindhoven University
of Technology, 5600 MB Eindhoven, The Netherlands}
\author{S. M. Frolov}
\affiliation{Kavli Institute of Nanoscience, Delft University of
Technology, 2600 GA Delft, The Netherlands}

\author{L. P. Kouwenhoven}
\affiliation{Kavli Institute of Nanoscience, Delft University of
Technology, 2600 GA Delft, The Netherlands}

\date{\today}
\begin{abstract}

Double quantum dot in the few-electron regime is achieved using local gating in an InSb nanowire. 
The spectrum of two-electron eigenstates is investigated using electric dipole spin resonance. Singlet-triplet 
level repulsion caused by spin-orbit interaction is observed. The size and the anisotropy of singlet-triplet 
repulsion are used to determine the magnitude and the orientation of the spin-orbit effective field in an 
InSb nanowire double dot. The obtained results are confirmed using spin blockade leakage current anisotropy 
and transport spectroscopy of individual quantum dots.

\end{abstract}

\maketitle

The spin-orbit interaction (SOI) describes coupling between the motion
of an electron and its spin. In one dimension, where electrons can move
only to the left or to the right, the SOI couples this left or right
motion to either spin-up or spin-down. An extreme situation occurs in
what is called a helical liquid \cite{Streda2003} where, in the
presence of magnetic field, all spin-up electrons move to the left and
all spin-down electrons to the right. As proposed recently \cite{Lutchyn2010,Oreg2010}, a
helical liquid in proximity to a superconductor can generate Majorana
fermions \cite{Kitaev2001}.  The search for Majorana fermions in 1D
conductors is focused on finding the best material in terms of a strong
spin-orbit interaction and large Land\' e \textit{g}-factors. The latter
is required for a helical liquid to exist at magnetic fields that do
not suppress superconductivity. High \textit{g}-factors of the order
50, strong SOI and the ability to induce superconductivity put forward InSb nanowires \cite{Nilsson2009, Nilsson2011} as a natural platform for the realization of 1D topological states. 

The SOI can be expressed as an effective magnetic field
${{\vec{{B}}}_{{\normalsubformula{\text{SO}}}}}$ that depends on the
electron momentum. An electron moving through the wire undergoes spin
precession around  ${{\vec{{B}}}_{{\normalsubformula{\text{SO}}}}}$
with a $\pi $ rotation over a distance
\textit{l}\textit{\textsubscript{SO }}called the spin-orbit length (see
Fig. 1(a)). The length \textit{l}\textit{\textsubscript{SO}} is a direct
measure of the SOI strength: a stronger SOI results in a shorter
\textit{l}\textit{\textsubscript{SO}}. In this letter, we use spin spectra of single
electrons in quantum dots \cite{Schreiber2010} to extract
\textit{l}\textit{\textsubscript{SO}} and the
direction of ${{\vec{{B}}}_{{\normalsubformula{\text{SO}}}}}$. In
quantum dots, the SOI hybridizes states with different
spin \cite{Nilsson2009,Pfund2007a,Fasth2007}. For a single electron, the
SOI-hybridized spin-up and spin-down states form a spin-orbit qubit \cite{Nadj-Perge2010a,Nowack2007}.
For two electrons SOI hybridization induces
level repulsion between singlet and triplet states. The resulting
level-repulsion gap between the well-defined qubit states can be used
to measure the SOI: the gap size is determined by
\textit{l}\textit{\textsubscript{SO}} \cite{Nilsson2009,Pfund2007a,Fasth2007} and the gap anisotropy indicates the
direction of $\vec{B}_{\text{SO}}$
\cite{Takahashi2010PRL, KanaiY2011,Golovach2008}. 

Double quantum dots in InSb nanowires are defined by local gating (Figs.
1(b),1(c)). A finite voltage is applied across the source and drain
electrodes; and the current through the nanowire is measured. Five
gates underneath the wire create the confinement potential and control
the electron number on the two dots \cite{Fasth2007,Nadj-Perge2010}. We focus
on the (1,1) charge configuration (Fig. 1(d)), in which both the left and
the right dot contain exactly one electron, each of them representing a
qubit \cite{Nadj-Perge2010a,Nowack2007,Pioro-Ladriere2008,Petta2005a,Koppens2006}.

\begin{figure}[htb]
\centering
\includegraphics[width=0.5\textwidth]{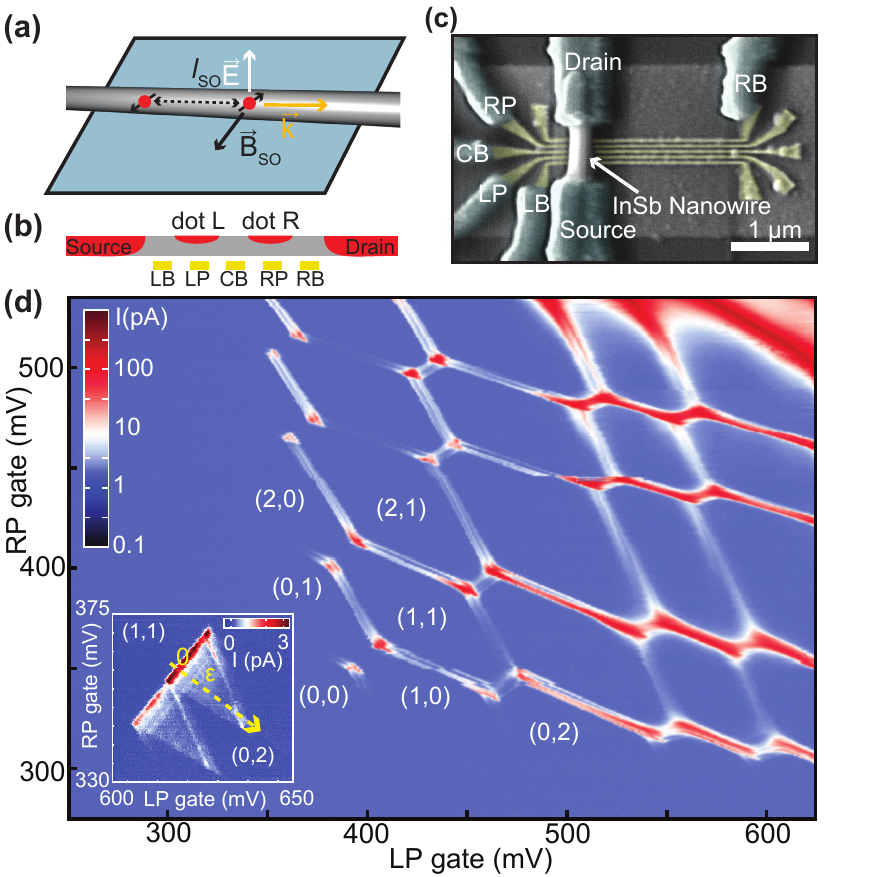}
\caption{ (color online) (a), An electron moving with momentum  ${\vec{{k}}}$ through
the wire experiences a spin-orbit field 
${{\vec{{B}}}_{{\normalsubformula{\text{SO}}}}}$ which rotates the spin
by $\pi $ after a distance \textit{l}\textit{\textsubscript{SO}}.
Vector  ${\vec{{E}}}$ indicates likely direction of the electric field.
In the case of spin-orbit coupling due to structural inversion
asymmetry, ${{\vec{{B}}}_{{\normalsubformula{\text{SO}}}}\propto
\vec{{E}}\times {\vec{{k}}}}$ \cite{Winkler2003}. (b),
 Schematic of a double quantum dot in an InSb nanowire. Red color
indicates regions of the nanowire which are not depleted by gates.
Gates LB, CB, and RB define the left, central and right barriers. Gates
LP and RP are the left and right plungers used to control the electron
number on each dot. (c), Scanning electron microscopy of a
nanowire device similar to the one used in the measurements. (d),
Charge stability diagram of the double dot for source-drain voltage
\textit{V}\textit{\textsubscript{sd}}= 1 mV. Typical
charging energy is 10 meV. Numbers in brackets correspond to the charge
occupation on the left and the right dots. \ The inset shows the charge
stability diagram near the (1,1) ${\rightarrow}$ (0,2) charge
transition for
\textit{V}\textit{\textsubscript{sd}}= 5 mV. The
detuning axis $\varepsilon $ is indicated by the dashed arrow. \ }

\label{fig:figure1}
\end{figure}

The qubit eigenstates are described by the Kramers spin-orbit doublet
${\Uparrow}$ and ${\Downarrow}$. These two states are superpositions of
spin-up and spin-down, and of several of the lowest orbital states
\cite{Flindt2006}. Similar to the case of pure spin states, a
magnetic field \textit{B} induces a Zeeman splitting
\textit{E}\textit{\textsubscript{Z }}\textit{=
g}\textit{${\mu}$}\textit{\textsubscript{B}}\textit{B} between the
Kramers doublets, where \textit{g} is the effective Land\' e
\textit{g}{}-factor for a given direction of  $\vec{B}$, and
$\mu_B$ is the Bohr
magneton. The two qubits in the (1,1) configuration can either form a
Kramers singlet state S(1,1) or one of the three triplets
T\textsubscript{+}(1,1), T\textsubscript{0}(1,1) and
T\textsubscript{{}-}(1,1). The states of the qubits are prepared using
Pauli spin blockade \cite{Nadj-Perge2010a,Nowack2007,Petta2005a,Koppens2006,Ono2002a} (Fig. 2(a)), which
relies on the tunneling process from (1,1) to the (0,2) spin singlet
S(0,2) (note that T(0,2) state is at 5 meV \ above S(0,2) and therefore
inaccessible for \textit{B} = 0). When the two electrons form a triplet
state, tunneling of the left electron to the right dot is prohibited by
selection rules. This absence of tunneling initializes the qubits in
the so-called blocked (1,1) state and thereby suppresses the current of
electrons passing through the double dot. Leakage current can occur due
to hybridization of T(1,1) states with S(0,2) induced by SOI and by
spin mixing between T(1,1) and S(1,1) due to hyperfine interaction
\cite{Nadj-Perge2010,Koppens2005, Pfund2007a, Danon2009a}.

\begin{figure}[h!]
\centering
\includegraphics[width=0.5\textwidth]{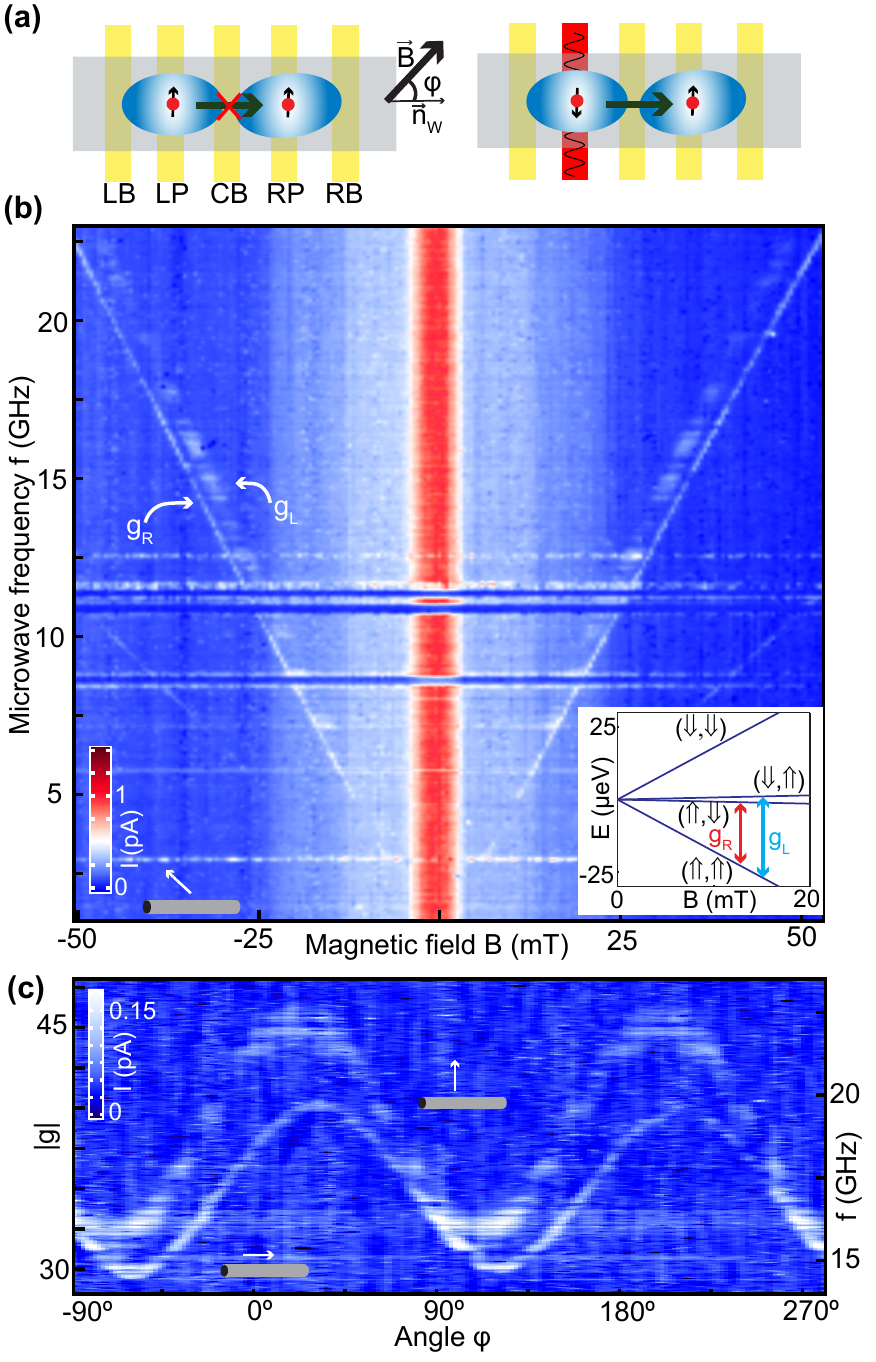}
\caption{(color online) (a), Left: blocked parallel configuration.
\ \textit{$\varphi $} is the angle between nanowire axis 
${{\vec{{n}}}_{{W}}}$ and  ${\vec{{B}}}$. Right: microwaves applied to
LP gate (red) induce EDSR. Tunneling to the right dot is allowed when
the left qubit is rotated to antiparallel configuration. (b) V-shaped EDSR resonances with slopes providing
\textit{g}\textit{\textsubscript{L}} and
\textit{g}\textit{\textsubscript{R}} for \textit{$\varphi $} =130 $^\circ$ and
\textit{V}\textit{\textsubscript{sd }}= 8 mV. Larger
\textit{g}{}-factor was assigned to the larger dot, i.e to the dot with
smaller orbital energy (orbital energy is 5 meV for the left dot and
7.5 meV for the right dot). V-shaped lines with half the slope are
two-photon transitions. Enhanced current around \textit{B}= 0 is due
to spin mixing in the absence of microwaves (see \cite{supp} section S2). Resonances at constant \textit{f} are due to
photon-assisted tunneling enhanced by cavity modes. (At each frequency
the maximum current is normalized to 1pA and a constant offset is
subtracted for clarity.) Inset shows energy spectrum of weakly coupled
double dots with arrows illustrating the observed transitions.
(c), Current versus \ \textit{f} \ and \textit{$\varphi $} for
\textit{B} = 35 mT. \ Vertical axis on the left is rescaled to
\textit{g=hf/}\textit{${\mu}$}\textit{\textsubscript{B}}\textit{B}. (At
each field a constant current offset is subtracted for clarity.) White arrows over grey cylinders indicate B-field orientation with respect to nanowire in panels (b) and (c).}
\label{fig:figure2}
\end{figure}

Transitions between qubit states are induced by a.c. electric fields via
electric dipole spin resonance (EDSR) \cite{Nadj-Perge2010a,Nowack2007,Pioro-Ladriere2008,Bell1962,Golovach2006,Laird2007}.
Voltages at microwave frequencies are applied to the left plunger (LP)
gate (Fig. 2(a)). The oscillating electric field wiggles the electronic
orbits. This periodic motion results, via SOI, in a rotation of the
spin \cite{Nadj-Perge2010a,Nowack2007}. When the microwave frequency is on
resonance with the double dot level transitions, EDSR can assist in
overcoming spin blockade thereby increasing the current through the
double dot. We map out this current increase as a function of microwave
frequency \textit{f }and ${\vec{{B}}}$ (Fig. 2(b)).

For weak interdot tunnel coupling the spectrum is determined by the
energies of individual qubits. At \textit{B=0} all four states are
degenerate and non-blocked due to fast decay to singlet state induced
by hyperfine interaction \cite{Koppens2005}. At finite \textit{B,}
parallel configurations (${\Uparrow}$, ${\Uparrow}$)=
T\textsubscript{+}(1,1) \ and \ (${\Downarrow}$, ${\Downarrow}$) =
T\textsubscript{{}-}(1,1) split in energy and become blocked, while the
other two configurations (${\Downarrow}$, ${\Uparrow}$) and
(${\Uparrow}$, ${\Downarrow}$) remain non-blocked. EDSR induces
transitions between {\textquoteleft}parallel{\textquoteright} and
{\textquoteleft}anti parallel{\textquoteright} configurations,
resulting in an on-resonance current as observed in Fig. 2(b). The slopes
of the two {\textquotedblleft}V{\textquotedblright} shaped resonances
determine the \textit{g}{}-factors of the right and left dots,
{\textbar}\textit{g}\textit{\textsubscript{R}}{\textbar} =
29.7{\textpm}0.2 and
{\textbar}\textit{g}\textit{\textsubscript{L}}{\textbar} =
32.2{\textpm}0.2 for this plot. Moreover, the g-factors of both dots
are highly anisotropic as revealed by the EDSR spectroscopy for
different field orientations (Fig. 2(c)). The observed anisotropy is
likely determined by the details of confinement \cite{Pryor2006,Schroer2011}
since the \textit{g}{}-factor in bulk zincblende InSb is expected to be
isotropic. 
\begin{figure}[htb]
\centering
\includegraphics[width=0.5\textwidth]{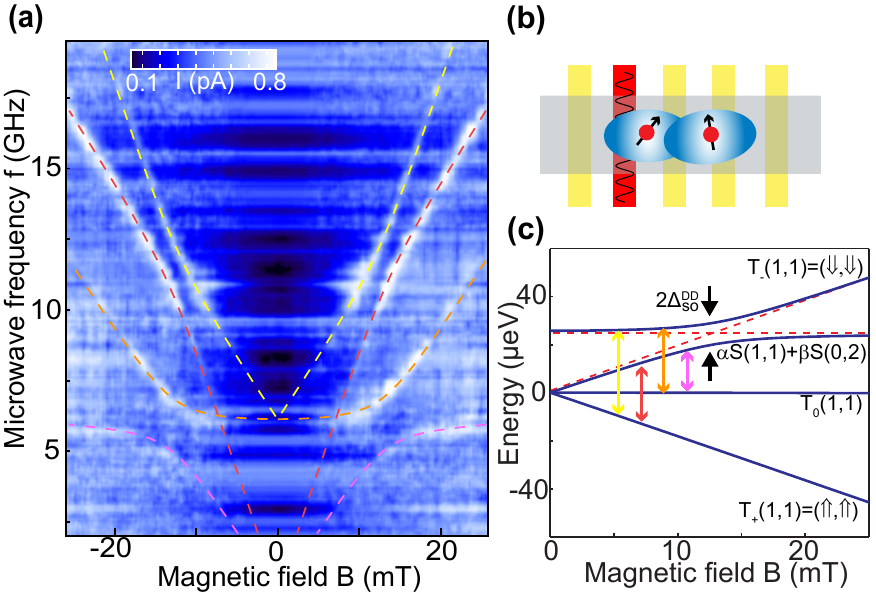}

\caption{(color online) (a), Current, in color, versus \textit{f} and
\textit{B} for detuning ${\varepsilon}$ ${\approx}$ 0.5 meV
(\textit{V}\textit{\textsubscript{sd }}=-5 mV). Dashed lines are fits
to a model described in the \cite{supp} section S4. Line
colors match transitions indicated in panel (c). (At each frequency a current offset is subtracted for
clarity.) (b), Diagram illustrating a strongly coupled double
quantum dot realized by applying a more positive voltage to the
\ central gate. (c), Energy diagram deduced from (a) and
used to extract the S-T spin-orbit gap  ${\Delta
_{{\normalsubformula{\text{SO}}}}^{{\normalsubformula{\text{DD}}}}}$.
Arrows indicate transitions observed in (a). In the absence of
coupling, the triplet and the singlet state would simply cross as
indicated by dashed lines.}
\label{fig:figure3}
\end{figure}

When we increase the interdot tunneling (Fig. 3(b)), the (1,1) states
hybridize with S(0,2) resulting in level repulsion between spectral
lines. In the absence of SOI, only states with the same spin can
hybridize \textit{e.g.} S(1,1) with S(0,2). SOI, however, also enables
hybridization between the singlets and the
triplets \cite{Schreiber2010,Fasth2007,Pfund2007, Danon2009a} (Fig. 3; see also Fig. 4(e)).
All observed transitions in Fig. 3(a) can be identified using a simple
model which takes into account the hybridization between the (1,1)
triplets and S(0,2) (see \cite{supp}  section S4). The
four avoided crossings observed in Fig. 3(a) correspond to the same
double dot spin-orbit gap  ${\Delta
_{{\normalsubformula{\text{SO}}}}^{{\normalsubformula{\text{DD}}}}}$
between T\textsubscript{{}-}(1,1) and the singlet, as illustrated in
Fig. 3(c). The quantitative comparison with the model allows us to
estimate the spin-orbit length
\textit{l}\textit{\textsubscript{SO}}\textit{ }= 230 {\textpm} 40 nm
(see \cite{supp} section S5). 

\begin{figure}[h!]
\centering
\includegraphics[width=0.5\textwidth]{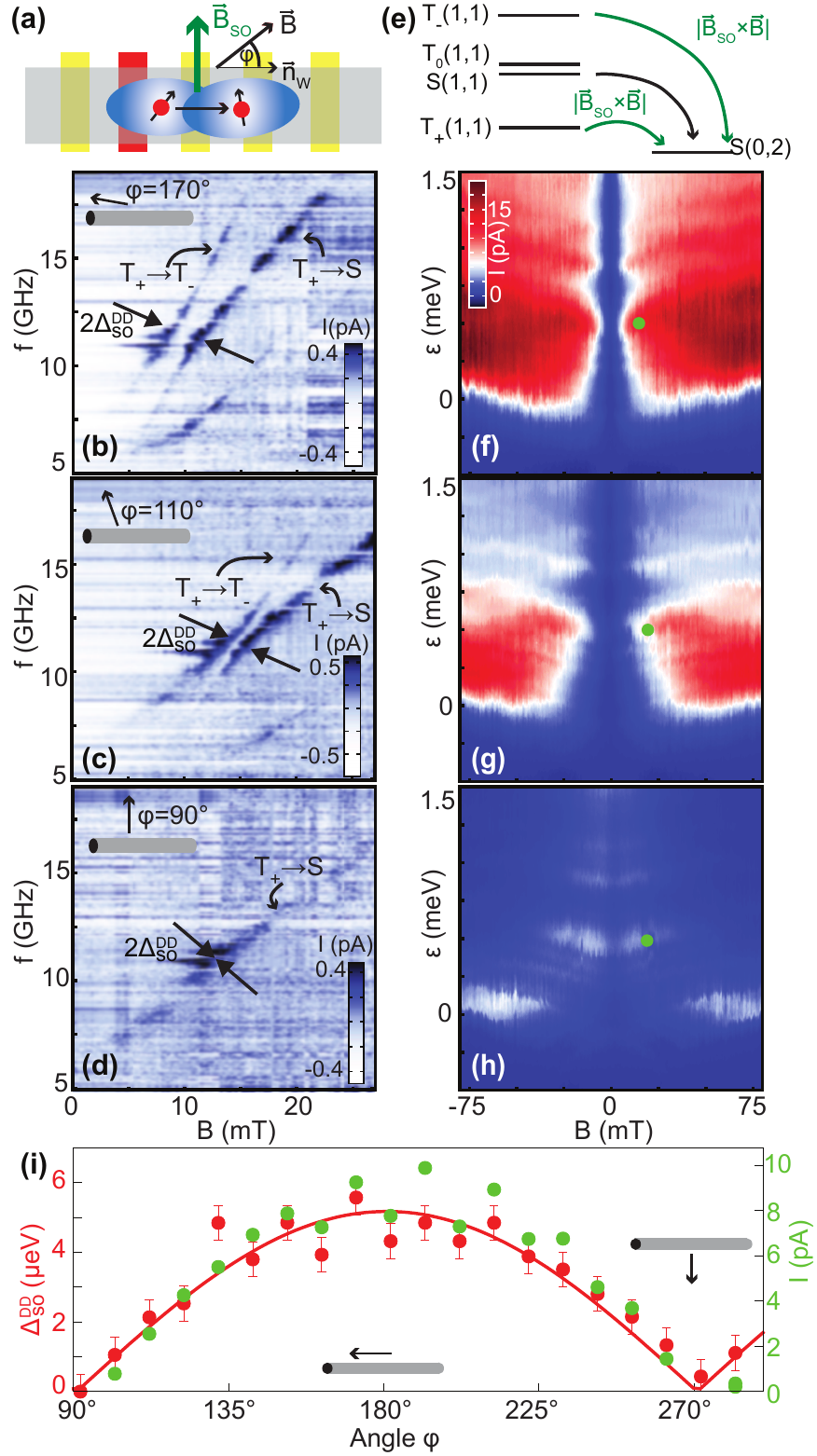}
\caption{ (color online) a, As the left electron tunnels to the right it experiences a field
${{\vec{{B}}}_{{\normalsubformula{\text{SO}}}}}$. (b-d) The avoided crossing in the EDSR spectrum as in Fig.
3(a) for three directions of  ${\vec{{B}}}$: \textit{$\varphi $} =170$^\circ$;
\textit{$\varphi $} = 110$^\circ$; and \textit{$\varphi $} = 90$^\circ$
(\textit{V}\textit{\textsubscript{sd}}=-5 mV). (At each magnetic field
an offset is subtracted for clarity.) (e), Transitions
between (1,1) states and S(0,2) at finite \textit{B}. \ The two singlet
states are hybridized due to tunnel coupling. T\textsubscript{+}(1,1)
and T\textsubscript{{}-}(1,1) are coupled to S(0,2) due to 
${{\vec{{B}}}_{{\normalsubformula{\text{SO}}}}}$. This SOI induced
coupling scales as 
${|{\vec{{B}}}_{{\normalsubformula{\text{SO}}}}\times {\vec{{B}}}|}$ for 
small ${\vec{{B}}}$ \cite{Danon2009a}. (f-h)\textbf{\textit{,}}\textbf{ \ }\textit{I} versus
\textit{$\varepsilon $} and \textit{B} for the same orientations of 
${\vec{{B}}}$ as in (b-d) with microwaves off.
(i), Extracted values of  ${\Delta
_{{\normalsubformula{\text{SO}}}}^{{\normalsubformula{\text{DD}}}}}$\textsubscript{
}\ (see \cite{supp} section S6) and $I$ at \textit{B}
= 20 mT and \textit{$\varepsilon $} = 0.5 meV (green dot in panels
(f-h)) as a function of \textit{$\varphi
$}. Solid line is a fit to  ${\Delta
_{{\normalsubformula{\text{SO}}}}^{{\normalsubformula{\text{DD}}}}}$=
${\Delta _{{\normalsubformula{\text{SO}}}}}${\textbar}cos\textit{
($\varphi $- $\varphi $}\textit{\textsubscript{0}}\textit{)}{\textbar}
with  ${\Delta _{{\normalsubformula{\text{SO}}}}}$=5.2{\textpm}0.3 $\mu
$eV and \textit{$\varphi $}\textit{\textsubscript{0}} = 1$^\circ${\textpm}5$^\circ$ .
The error bars are determined by the width of EDSR resonance. }
\label{fig:figure4}
\end{figure}

The observed singlet-triplet gap is highly anisotropic (Fig. 4). The gap
 ${\Delta
_{{\normalsubformula{\text{SO}}}}^{{\normalsubformula{\text{DD}}}}}$ is
largest when  ${\vec{{B}}}$ is parallel to the nanowire axis
${{\vec{{n}}}_{{W}}}$:  ${\Delta
_{{\normalsubformula{\text{SO}}}}^{{\normalsubformula{\text{DD}}}}}$
shrinks as the direction of  ${\vec{{B}}}$ is rotated in the sample
plane (Fig. 4(b) and 4(c)). Finally for ${\vec{{B}}\perp
{\vec{{n}}}_{{W}}}$ the gap disappears (Fig. 4(d)). For this orientation
the resonance line corresponding to the T\textsubscript{+}(1,1) \ to
singlet transition becomes straight indicating the absence of level
repulsion between T\textsubscript{{}-}(1,1) and singlet. In addition,
the visibility of the T\textsubscript{+}(1,1) $\rightarrow $
T\textsubscript{{}-}(1,1) transition vanishes, suggesting that both
T\textsubscript{+}(1,1) and T\textsubscript{{}-}(1,1) states are
completely blocked for this field orientation.

The observed anisotropy of  ${\Delta
_{{\normalsubformula{\text{SO}}}}^{{\normalsubformula{\text{DD}}}}}$
confirms the spin-orbit origin of the singlet-triplet level repulsion
(see also \cite{supp} section S3). The gap ${\Delta
_{{\normalsubformula{\text{SO}}}}^{{\normalsubformula{\text{DD}}}}}$ is
expected to be proportional to 
${|{\vec{{B}}}_{{\normalsubformula{\text{SO}}}}\times (\vec{{B}}/B)|}$
\cite{Danon2009a, Nowak2011, Stepanenko2011}. \ When the two fields are aligned, singlet and
triplet states cannot mix and therefore the spin-orbit gap closes (Fig.
4(d)). From the observed anisotropy we conclude that 
${{\vec{{B}}}_{{\normalsubformula{\text{SO}}}}}$ points perpendicular
to the nanowire and is parallel to the substrate plane (Figs. 4(i) and
4(a)).

The knowledge of  ${{\vec{{B}}}_{{\normalsubformula{\text{SO}}}}}$
orientation provides a substantial increase in the fidelity of the
initialization and readout of spin-orbit qubits \cite{Nadj-Perge2010a}.
The fidelity is presently limited due to unwanted transitions from
T\textsubscript{+}(1,1) and T\textsubscript{{}-}(1,1) to S(0,2) induced
by SOI. When  ${\vec{{B}}}$ and 
${{\vec{{B}}}_{{\normalsubformula{\text{SO}}}}}$ are misaligned,
T\textsubscript{+}(1,1) and T\textsubscript{{}-}(1,1) are coupled to
S(0,2) (Fig. 4(e)) \cite{Danon2009a}. The unwanted transitions are
manifest in the d.c. current through the double dot at finite magnetic
fields (Figs. 4(f), 4(g), 4(h)) \cite{Nadj-Perge2010,Pfund2007}.  For an ideal
readout and initialization no current flows after either
T\textsubscript{+}(1,1) or T\textsubscript{{}-}(1,1) state is occupied.
When  ${\vec{{B}}}$ is aligned with
${{\vec{{B}}}_{{\normalsubformula{\text{SO}}}}}$,
T\textsubscript{+}(1,1) and T\textsubscript{{}-}(1,1) become decoupled
from S(0,2) and d.c. current is expected to vanish. This dramatic
suppression of d.c. current is observed for  ${\vec{{B}}\perp
{\vec{{n}}}_{{W}}}$(Fig. 4(h)). Importantly, both  ${\Delta
_{{\normalsubformula{\text{SO}}}}^{{\normalsubformula{\text{DD}}}}}$
and \textit{I} show almost identical angle dependence further
confirming that the singlet-triplet hybridization due to SOI is absent
when  ${\vec{{B}}}${\textbar}{\textbar}
${{\vec{{B}}}_{{\normalsubformula{\text{SO}}}}}$ (Fig. 4(i)). 

Given the direction of  ${{\vec{{B}}}_{{\normalsubformula{\text{SO}}}}}$
we can analyze the origin of the spin-orbit interaction in InSb
nanowires. The field  ${{\vec{{B}}}_{{\normalsubformula{\text{SO}}}}}$
depends on the electron momentum  ${\vec{{k}}}$. In a simple physical
picture, during the interdot tunneling, the momentum  ${\vec{{k}}}$ is
along the nanowire, which is grown in the [111] crystallographic
direction. In zincblende InSb the spin-orbit interaction has two
contributions, the bulk-inversion asymmetry term (BIA) and the
structure-inversion asymmetry term (SIA). However, for  ${\vec{{k}}}$
{\textbar}{\textbar} [111] the BIA term is expected to vanish
\cite{Winkler2003}, and therefore the SIA contribution should
dominate. The field  ${{\vec{{B}}}_{{\normalsubformula{\text{SO}}}}}$
due to SIA is orthogonal to both the momentum and the external electric
field (Fig. 1(c)). The electric field is likely perpendicular to the
substrate since the symmetry of confinement in the nanowire is broken
by the substrate dielectric and voltages on the gates. \ Therefore the
direction  ${{\vec{{B}}}_{{\normalsubformula{\text{SO}}}}\perp
{\vec{{n}}}_{{W}}}$ and in the substrate plane is consistent with the
SIA spin-orbit interaction.

We compare the results obtained from EDSR spectroscopy with the spectrum
of (0,2) states (Fig. 5(a)) \cite{Nilsson2009,Pfund2007,Fasth2007}. The SOI
hybridization of S(0,2) and T\textsubscript{+}(0,2) states leads to a
single dot spin-orbit gap ${\Delta
_{{\normalsubformula{\text{SO}}}}^{{\normalsubformula{\text{SD}}}}}$.
Since the energies of the (0,2) states are too large to be accessed
with microwaves (${\Delta}$\textsubscript{ST }${\approx}$ 5 meV at
\textit{B}=0), we use the lowest energy T\textsubscript{+}(1,1) level
as a probe of the (0,2) spectrum. By changing detuning we move
T\textsubscript{+}(1,1) with respect to the (0, 2) levels. \ When
T\textsubscript{+}(1,1) is aligned with either S(0,2) or
T\textsubscript{+}(0,2), an increase in d.c. current is observed (Fig.
5(b)) \cite{Pfund2007a}. The level repulsion between
T\textsubscript{+}(0,2) and S(0,2) \ is observed at \textit{B
}${\approx}$ 2T (Fig. 5(c)). The single dot gap is also strongly
anisotropic reaching the smallest value for  ${\vec{{B}}\perp
{\vec{{n}}}_{{W}}}$ (Figs. 5(d), 5(e) and 5(f)). The spin-orbit length
\textit{l}\textit{\textsubscript{SO}} = 310 {\textpm} 50 nm estimated
from  ${\Delta
_{{\normalsubformula{\text{SO}}}}^{{\normalsubformula{\text{SD}}}}}$ is
in agreement with the value obtained using EDSR.

\begin{figure}[h!]
\centering
\includegraphics[width=0.5\textwidth]{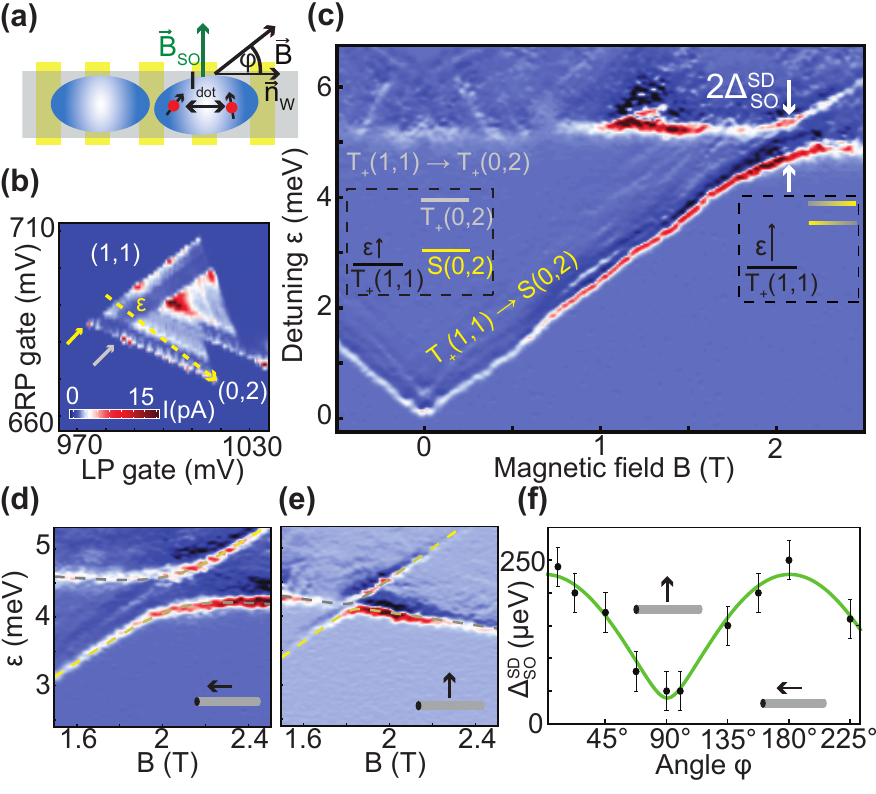}
\caption {(color online) (a), Two electrons in the right quantum dot. The separation of the two
electrons in the triplet state is of the order of the dot size.
(b), Charge stability diagram close to (1,1) $\rightarrow $
(0,2) transition at \textit{B=} 1.4 T, for
\textit{V}\textit{\textsubscript{sd}}= 7 mV and 
${\vec{{B}}||{\vec{{n}}}_{{W}}}$. Transitions T\textsubscript{+}(1,1)
$\rightarrow $ S(0,2) and T\textsubscript{+}(1,1) $\rightarrow $
T\textsubscript{+}(0,2) are indicated by yellow and gray arrows.
(c), Resonances corresponding to T\textsubscript{+}(1,1)
$\rightarrow $ S(0,2) and T\textsubscript{+}(1,1) $\rightarrow $
T\textsubscript{+}(0,2) as a function of \ \textit{B} for
\textit{$\varphi $=}180$^\circ$. Colors from dark blue (low) to red (high) in
panes (c), (d) and (e) indicate values of
\textit{dI/}d$\varepsilon $ in arbitrary units. (d) and 
(e), Avoided crossing for \textit{$\varphi $=}180$^\circ$ and \textit{$\varphi
$=}90$^\circ$. The dashed lines are fits to the model from Ref. \cite{Fasth2007}.
(f), The gap ${\Delta
_{{\normalsubformula{\text{SO}}}}^{{\normalsubformula{\text{SD}}}}}$ as
a function of \textit{$\varphi $}. Solid line is a fit to  ${\Delta
_{{\normalsubformula{\text{SO}}}}^{{\normalsubformula{\text{SD}}}}=\Delta
'_{{\normalsubformula{\text{SO}}}}\sqrt{\text{cos}^{{2}}(\varphi
-\varphi _{{0}})\text{cos}^{{2}}\theta +\text{sin}^{{2}}\theta }}$ with
\  ${\Delta '_{{\normalsubformula{\text{SO}}}}}$= 230 {\textpm} 10 $\mu
$eV, \textit{$\varphi $}\textit{\textsubscript{0}} = 2$^\circ$ {\textpm} 5$^\circ$
and \textit{$\theta $=}10$^\circ$\textit{ }{\textpm} 3$^\circ$. The error bars are
determined by average linewidth corresponding to
T\textsubscript{+}(1,1) $\rightarrow $ S(0,2) and
T\textsubscript{+}(1,1) $\rightarrow $ T\textsubscript{+}(0,2)
transitions. Note that the anisotropy of  ${\Delta
_{{\normalsubformula{\text{SO}}}}^{{\normalsubformula{\text{SD}}}}}$
depends on the relative positions of the two electrons in the right dot
which may be different from nanowire axis. Out-of-plane
${{\vec{{B}}}_{{\normalsubformula{\text{SO}}}}}$ angle \textit{$\theta
$} therefore may be non-zero due to confinement details of the right
quantum dot. Measurements at the (1,1) $\rightarrow $ (2,0) transition
yielded the same in-plane anisotropy for the left dot (data not
shown).}
\label{fig:figure5}
\end{figure}

Recent proposals for experimental detection of Majorana bound states in hybrid nanowire-superconductor
devices require wires with strong spin-orbit coupling \cite{Lutchyn2010,Oreg2010}. Besides InSb, indium arsenide (InAs) and p-type silicon/germanium (Si/Ge) nanowires \cite{kloeffel2011} are among most promising material systems for this purpose. Majorana states are expected to appear at the boundaries of 
the topological superconducting phase. The topological phase is predicted to occur if: (i)
\textit{E}\textit{\textsubscript{Z}}\textit{ {\textgreater}} $\Delta
$ and (ii)
\textit{E}\textit{\textsubscript{top}}, $\Delta $ {\textgreater}
\textit{T}. \ Here $\Delta $ is the superconducting gap,
\textit{E}\textit{\textsubscript{top}} is the gap of the topological
phase and \textit{T} is the temperature. Due to large \textit{g}-factors in InSb nanowires first requirement is satisfied at 
low magnetic fields even if large gap superconductors such as niobium are used ($\Delta $ $\sim$ 5K). This is a clear advantage since low magnetic fields are preferential in order not to suppress superconductivity. The size of the topological gap 
${E_{{\normalsubformula{\text{top}}}}\approx
2\sqrt{E_{{\normalsubformula{\text{SO}}}}\Delta }}$ is determined by
the bulk SOI splitting \textit{E}\textit{\textsubscript{SO}} =  ${\hbar
^{{2}}/(2m_{{e}}^{{*}}l_{{\normalsubformula{\text{SO}}}}^{{2}})}$
\cite{Streda2003}. Here  ${\hbar }$ is the Planck constant and 
${m_{{e}}^{{*}}\approx 0\text{.}\text{015}\,m_{{e}}}$ is the effective
electron mass (${m_{{e}}}$ is the electron mass). We can estimate
\textit{E}\textit{\textsubscript{SO}} ${\approx}$ 0.5 K and
\textit{E}\textit{\textsubscript{top}} ${\approx}$ 3
K for the case of ballistic one-dimensional transport. While \textit{E}\textit{\textsubscript{SO}} is expected to be an order of magnitude larger for p-type Si/Ge wires  \cite{kloeffel2011} the  \textit{E}\textit{\textsubscript{SO}} ${\approx}$ 0.1-0.3 K is similar for InAs wurtzite nanowires \cite{supp} ($m_e^{*}\approx 0.042$-$0.06\, m_e$ for wurtzite InAs \cite{pryor2010}). Note however that besides strength of SOI experimental details such as quality of semiconductor-superconductor interface as well as disorder may in the end determine the most promising material system.  Finally we note that the anisotropy measurements (Fig. 4 and 5) suggest the orientation ${\vec{{B}}||{\vec{{n}}}_{{W}}}$ to be optimal for observing Majorana states since the maximum mixing of the SOI-split
bands occurs for  ${\vec{{B}}\perp {\vec{{B}}}_{{\normalsubformula{\text{SO}}}}}$ and the
superconductivity is suppressed least when  ${\vec{{B}}}$ is in the
substrate plane.

We would like to thank J. Danon, Y. Nazarov, M. Rudner, D. Loss, F. Hassler and J. van Tilburg for discussions and help. We acknowledge help with the measurement software from R. Heeres and P. de Groot. This work has been supported by ERC, NWO/FOM Netherlands Organization for Scientific Research and through the DARPA program QUEST.

\end{document}